\documentclass{article}

\usepackage{graphicx}
\usepackage[hyphens]{url}
\usepackage{multicol,multirow}
\usepackage{amsmath,amssymb,amsfonts}
\usepackage{mathrsfs}
\usepackage{amsthm}
\usepackage{rotating}
\usepackage{appendix}
\usepackage[a4paper, margin=1in]{geometry}
\usepackage[numbers]{natbib}
\usepackage{ifpdf}
\usepackage[T1]{fontenc}
\usepackage{txfonts}
\usepackage{textcomp}
\usepackage{xcolor}
\usepackage{soul}
\usepackage{lipsum}
\usepackage[colorlinks,allcolors=blue]{hyperref}
\usepackage{cleveref}
\usepackage{tikz}
\usepackage{indentfirst}
\tikzstyle{arrow} = [->,>=stealth]
\usetikzlibrary{shapes, positioning, automata}
\usepackage{float}
\usepackage{authblk}
\usepackage{hyperref}
\restylefloat{table}

\theoremstyle{definition}

\numberwithin{equation}{section}

\title{Revisiting Taylor and the Trinity Test}
\author[1]{Elizabeth Mone}
\author[2]{Pranay Seshadri}
\affil[1]{College of Sciences, Georgia Institute of Technology, Atlanta, Georgia}
\affil[2]{College of Engineering, Georgia Institute of Technology, Atlanta, Georgia}
\date{}

\begin{document}

\maketitle
\section*{Abstract}
The atomic bomb uses fission of heavy elements to produce a large amount of energy. It was designed and deployed during World War II by the United States military. The first test of an atomic bomb occurred in July of 1945 in New Mexico and was given the name Trinity; this test was not declassified until 1949. In that year Geoffrey Ingram Taylor released two papers, detailing his process in calculating the energy yield of the atomic bomb from pictures of the Trinity explosion alone. Many scientists made similar calculations concurrently, though Taylor is often accredited with them. Since then many scientists have also attempted a to calculate a yield through various methods. This paper walks through these methods with a focus on Taylor's method---based on first principles---as well as redoing the calculations that he performed with modern tools. In this paper we make use of the state-of-the-art computer vision tools to find a more precise measurement of the blast radius, as well as using curve fitting and numerical integration methods. With more precise measurements we are able to follow in Taylor's footstep towards a more accurate approximation.

\tableofcontents

\section[Introduction]{Introduction}
In most introductory engineering fluid mechanics and thermodynamic courses, the work of Sir Geoffrey Ingram Taylor \citep{RN5, RN6} is presented as an important application of dimensional analysis: using the Buckingham-Pi theorem \citep{RN7} to estimate the power of an atomic bomb by studying a few images of the blast. Over the years there have been numerous surveys of Taylor's work \citep{RN1, RN2, RN3}, all emphasizing that Taylor \emph{did not} invoke the dimension reduction analysis he is often incorrectly credited with. Arguably his work was far more profound; starting with underlying equations and introducing a series of simplifications in concert with observations to work out a chosen quantity of interest. The approach Taylor takes is particularly salient and relevant for the broad (beyond fluids and thermodynamics) Data-Centric Engineering (DCE) audience, as it offers insights into studying data-driven physical processes. In this brief survey article, the work of Taylor is revisited and communicated using a series of illustrations, state-of-the-art computer vision software, and commentary---particularly on uncertainty in equations and data. The aim is to make Taylor's original papers more accessible, and in doing so highlight how much information regarding a problem can be obtained without resorting to computational simulations. The exposition here is distinct from the aforementioned surveys and is intended for the DCE readership that cuts across engineering, statistics, and machine learning. 

\section{Background}

\begin{figure}[h]
    \centering
    \includegraphics[width=\textwidth]{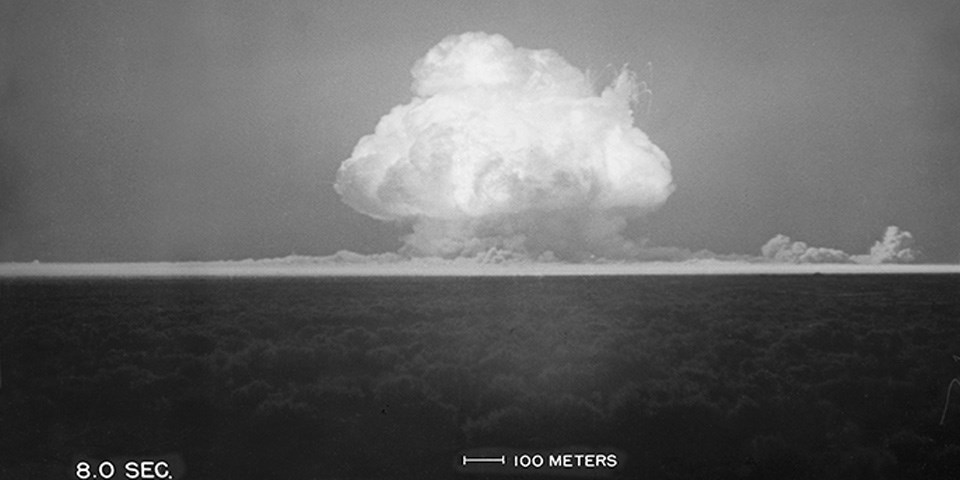}
    \caption{Picture of the Trinity Explosion \citep{RN13}}
    \label{fig:explosion}
\end{figure}

The Trinity explosion, shown in Figure~\ref{fig:explosion}, was the name for the first test of the atomic bomb that occurred on July $16^{th}$, 1945 in New Mexico, just a month before the bombs were dropped into Hiroshima and Nagasaki. The test bomb released a huge amount of energy causing radioactive fallout in the area and a blast that could be seen from space, reaching 40,000 feet in only 7 minutes \citep{RN9}. The atomic bomb would go on to end the war and have devastating consequences for the civilians of Japan, and raise profound questions on the very nature of our existence and the use of weapons of such cataclysmic destruction. Those involved in the Manhattan project\footnote{We note that our submission coincides with the release of Christopher Nolan's \emph{Oppenheimer}, that details the life of the chief scientist of the Manhattan project, Robert J. Oppenheimer.}, i.e., the name of the team that developed the atomic bomb, as well as others such as Taylor, spent much time actualizing the idea of a fission bomb and understanding the effects of such a device. 

There are multiple approaches one can take when calculating the energy of a blast, such as one from an atomic bomb. During the time of the Trinity test and World War II there were three main minds working on the problem. Taylor, the focus of this paper, is often cited with the calculation as he was the only one who gave proof of the validity of a point source model, as well as calculating the energy within a range of error. Two others, John von Neumann and Russian Leonid Sedov also performed similar calculations at the time, and actually achieved better accuracy than Taylor. Taylor's approach was to use partial differential forms of the equations for motion, continuity, and state as well as integral forms of energy to arrive at an approximate formula for the energy of the blast wave. Sedov and von Neumann followed similar approaches that began with dimensional analysis. The two then determined integral forms for energy and computed them analytically, though neither directly calculated energy. A more in-depth exploration of their calculations are provided in their own papers \citep{1946JApMM..10..241S, RN12} and in the work of Deakin \citep{RN2}. All three did not find exact values as the radiative energy was not taken into account \citep{RN2}. 

\begin{table}[h]
    \centering
    \caption{Summary of Trinity test  calculations.}
    \begin{tabular}{|c|c|}
        \hline
        Author & Energy (kilotons TNT) \\ \hline \hline
        Taylor \citep{RN6} &  16.8 \\ 
        Sedov, von Neumann \citep{RN2} & 16.9\\
        Deakin \citep{RN2} & 17.5 \\
        Hanson \citep{RN8} & 22.1 \\
        Truman \citep{RN15} & 20 \\
        Groves \citep{RN14} & 15-20 \\
        \hline
    \end{tabular}
    \label{tab:energy estimates}
\end{table}

Table~\ref{tab:energy estimates} shows the energy estimates of various authors including Taylor's estimate. Taylor's, Deakin's, Sedov's, and von Neumann's calculations are all under the assumption that the specific heat ratio $\gamma$ is constant and equal to $1.4$. Hanson's calculations---done more recently---use physical samples from the explosion to calculate the amount of plutonium and other elemental constituents of the bomb, from which the yield is determined \citep{RN8}. President Truman and General Groves, who was involved in the Manhattan project, released the values presented in the table once the information was declassified \citep{RN2, RN3}.

\subsection{Notation}
Table~\ref{tab:notation} contains the variables used both in this paper and by Taylor in \citep{RN5, RN6}, and will be used throughout this brief paper. 
\begin{table}[h]
    \centering
    \caption{Nomenclature.}
    \begin{tabular}{|c|c|c|c|}
        \hline
        Symbol & Quantity & Symbol & Quantity \\ \hline \hline
        $p$ & pressure & $p_0$ & stagnation pressure \\ 
        $\rho$ & density & $\rho_0$ & stagnation density \\
        $u$ & radial velocity & $t$ & time\\
        $E$ & energy & $\gamma$ & heat capacity ratio\\
        $R$ & shock radius & $r$ & radial position \\
        \hline
    \end{tabular}
    \label{tab:notation}
\end{table}

\subsection{Dimensional analysis approach}
Prior to detailing Taylor's approach, it will be useful to touch upon dimensional analysis with the Buckingham Pi theorem. This theorem states that with $N$ variables and $K$ fundamental units there are $P=\left( N-K \right)$ dimensionless quantities \citep{RN7}. By using this theorem we can identify the unit-less quantities that will enable us to calculate the energy of the explosion. To complete such an analysis the three scientists, Taylor, Sedov, and von Neumann, reduced the explosion to starting from a point source and expanding outwards in a spherical shock wave, rapidly releasing large amounts of energy. The use of a shock wave suggests that the state of the air can be expressed in two categories: internal shock wave properties and external conditions of the undisturbed atmosphere. Therefore, we can devise seven variables: $E$ the energy, $R$ the radius, $t$ the time, $p$ the static pressure of the wave, $p_0$ the pressure of the undisturbed atmosphere or stagnation pressure, $\rho$ the static density of the shock wave, and $\rho_0$ the undisturbed density of the atmosphere or stagnation density. All variable units, when broken down into their base units, contain three fundamental units: mass, length, and time. In terms of the Buckingham Pi theorem there are $N=7$ variables and $K=3$ fundamental quantities, therefore this problem contains $P=4$ dimensionless quantities \citep{RN2}. 

We wish to focus on a quantity that involves the target variable of energy, as well as easily measurable quantities which in this case are radius, time, and the density of stagnant air. To find the exact form of the dimensionless quantity we can use dimensional analysis to solve for energy. In SI units energy is recorded in Joules $(J)$ which is equivalent to a Newton-meter $(N\cdot m)$ or kilogram-meter-squared-per-second-squared $\left( kg\cdot m^2 / s^2 \right)$. Therefore, an equation for energy must contain units of mass $(kg)$, length $(m)$, and time $(s)$. Mass is not directly used within the context of the problem, instead density (units of $kg/m^3$) takes its place. In Taylor's paper he is interested in the time since the blast, the density of the undisturbed atmosphere (stagnation), and the radius of the shock wave. Thus, the equation for energy has the form 
\begin{equation}
    E = K \rho_0^a R^b t^c
\end{equation}
where $a, b, c, K$ are constant \citep{RN1}.

There are a few important points to note regarding the above. Firstly, energy is per second squared and since neither density nor radius contain units of time, the equation for energy must contain time, leading to the negative second power, or $c=-2$. Secondly, density is the only variable that contains units of mass, and since both energy and density contain mass only to the first power, then $a$ must equal 1. However, the inclusion of density adds an additional $m^{-3}$ to the equation that must be corrected in order to get $m^2$ in the final form. Therefore, to fix this we need a factor of $m^5$ which can be provided by the radius, meaning $b=5$ and we have that
\begin{equation} \label{eq:E}
    E=K\rho_0 R^5t^{-2},
\end{equation} 
where $K$ is a constant and proven to be a function of $\gamma$ by Taylor \citep{RN5}. Dimensional analysis is taken from \citep{RN1}. 

\subsection{Taylor's first principles approach}
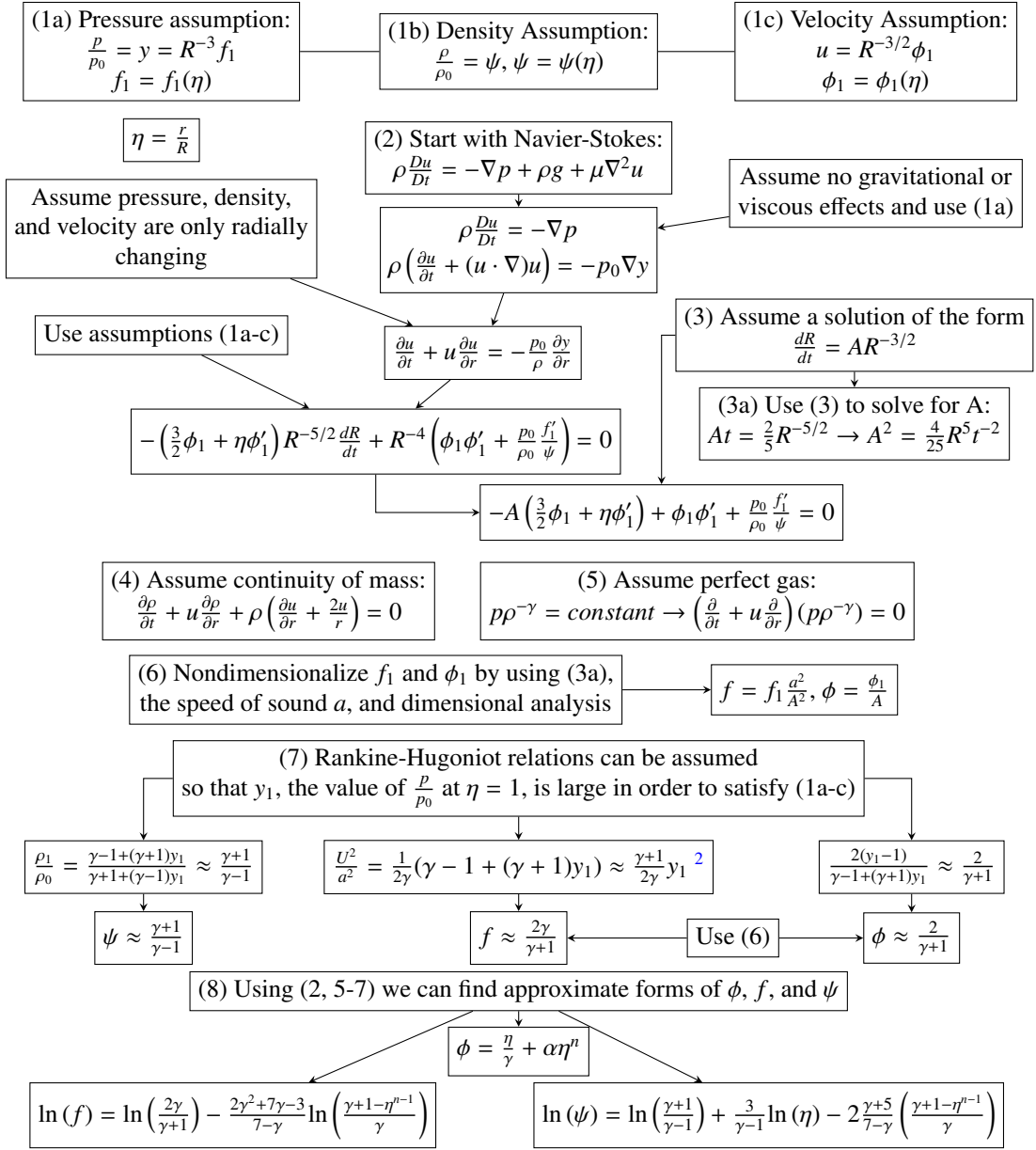
\begin{figure}[h]
\centering
\begin{tikzpicture}
    \node (p1a) [draw, align=center] at (-6,0) {(1a) Pressure assumption: \\ $\frac{p}{p_0}=y=R^{-3}f_1$ \\ $f_1=f_1(\eta)$};
    \node (p1b) [draw, align=center] at (-1,0) {(1b) Density Assumption: \\ $\frac{\rho}{\rho_0}=\psi$, $\psi=\psi(\eta)$};
    \draw (p1a) -- (p1b);
    \node (p1c) [draw, align=center] at (4,0) {(1c) Velocity Assumption: \\ $u=R^{-3/2}\phi_1$ \\ $\phi_1=\phi_1(\eta)$};
    \draw (p1b) -- (p1c);
    \node (p1d) [draw, align=center] at (-6, -1.25) {$\eta=\frac{r}{R}$};
    
    \node (p2) [draw, align=center] at (-1,-1.5) {(2) Start with Navier-Stokes: \\ $\rho\frac{Du}{Dt}=-\nabla p +\rho g +\mu\nabla^2u$};
    \node (p2a) [draw, align=center] at (-6, -2.5) {Assume pressure, density, \\ and velocity are only radially \\ changing};
    \node (p2b) [draw, align=center] at (4, -2) {Assume no gravitational or \\ viscous effects and use (1a)};
    \node (p2c) [draw, align=center] at (-1, -2.8) {$\rho \frac{Du}{Dt}=-\nabla p$ \\ $\rho\left(\frac{\partial u}{\partial t} + (u \cdot \nabla)u\right) = -p_0 \nabla y$};
    \draw[arrow] (p2) -- (p2c);
    \draw[arrow] (p2b) -- (p2c);
    \node (p2d) [draw, align=center] at (-1.5, -4.25) {$\frac{\partial u}{\partial t}+u
    \frac{\partial u}{\partial r} = -\frac{p_0}{\rho}\frac{\partial y}{\partial r}$};
    \draw[arrow] (p2a) -- (p2d);
    \draw[arrow] (p2c) -- (p2d);
    \node (p2e) [draw, align=center] at (-6, -4) {Use assumptions (1a-c)};
    
    \node (p3) [draw, align=center] at (3.7, -4) {(3) Assume a solution of the form \\ $\frac{dR}{dt}=AR^{-3/2}$};
    \node (p3a) [draw, align=center] at (3.7, -5.25) {(3a) Use (3) to solve for A: \\ $At=\frac{2}{5}R^{-5/2} \rightarrow A^2=\frac{4}{25}R^5t^{-2}$};
    \draw[arrow] (p3) -- (p3a);
    \node (p2f) [draw, align=center] at (-3, -5.5) {$-\left(\frac{3}{2}\phi_1 +\eta \phi_1'\right)R^{-5/2}\frac{dR}{dt} +R^{-4}\left(\phi_1\phi_1' + \frac{p_0}{\rho_0}\frac{f_1'}{\psi}\right)=0$};
    \draw[arrow] (p2d) -- (p2f);
    \draw[arrow] (p2e) -- (p2f);
    \node (p23) [draw, align=center] at (1,-6.5) {$-A\left(\frac{3}{2}\phi_1 +\eta \phi_1'\right) +\phi_1\phi_1' + \frac{p_0}{\rho_0}\frac{f_1'}{\psi}=0$};
    \draw[arrow] (p2f) |- (p23);
    \draw[arrow] (p3) -| (p23);
    
    \node (p4) [draw, align=center] at (-4.5, -7.75) {(4) Assume continuity of mass: \\ $\frac{\partial \rho}{\partial t}+ u\frac{\partial \rho}{\partial r} + \rho\left(\frac{\partial u}{\partial r}+\frac{2u}{r}\right)=0$};
    \node (p5) [draw, align=center] at (1.5, -7.75) {(5) Assume perfect gas: \\ $p\rho^{-\gamma}=constant \rightarrow \left(\frac{\partial}{\partial t} + u\frac{\partial}{\partial r}\right)\left(p\rho^{-\gamma}\right)=0$};

    \node (p6) [draw, align=center] at (-3, -9) {(6) Nondimensionalize $f_1$ and $\phi_1$ by using (3a), \\ the speed of sound $a$, and dimensional analysis};
    \node (p6a) [draw, align=center] at (3, -9) {$f=f_1\frac{a^2}{A^2}$, $\phi=\frac{\phi_1}{A}$};
    \draw[arrow] (p6) -- (p6a);
    
    \node (p7) [draw, align=center] at (-1, -10.25) {(7) Rankine-Hugoniot relations can be assumed \\ so that $y_1$, the value of $\frac{p}{p_0}$ at $\eta=1$, is large in order to satisfy (1a-c)};
    \node (p7a) [draw, align=center] at (-6.25, -11.5) {$\frac{\rho_1}{\rho_0}=\frac{\gamma-1+(\gamma+1)y_1}{\gamma +1+(\gamma-1)y_1} \approx \frac{\gamma+1}{\gamma-1}$};
    \draw[arrow] (p7) -| (p7a);
    \node (p7b) [draw, align=center] at (-1, -11.5) {$\frac{U^2}{a^2}=\frac{1}{2\gamma}(\gamma-1+(\gamma+1)y_1) \approx \frac{\gamma+1}{2\gamma}y_1$ \footnotemark};
    \draw[arrow] (p7) -- (p7b);
    \node (p7c) [draw, align=center] at (4.5, -11.5) {$\frac{2(y_1-1)}{\gamma-1+(\gamma+1)y_1} \approx \frac{2}{\gamma+1}$};
    \draw[arrow] (p7) -| (p7c);
    \node (p7d) [draw, align=center] at (-6.25, -12.5) {$\psi \approx \frac{\gamma+1}{\gamma-1}$};
    \draw[arrow] (p7a) -- (p7d);
    \node (p7e) [draw, align=center] at (-1, -12.5) {$f \approx \frac{2\gamma}{\gamma+1}$};
    \draw[arrow] (p7b) -- (p7e);
    \node (p7f) [draw, align=center] at (4.5, -12.5) {$\phi \approx \frac{2}{\gamma+1}$};
    \draw[arrow] (p7c) -- (p7f);
    \node (p7g) [draw, align=center] at (2, -12.5) {Use (6)};
    \draw[arrow] (p7g) -- (p7f);
    \draw[arrow] (p7g) -- (p7e);

    \node (p8) [draw, align=center] at (-1, -13.25) {(8) Using (2, 5-7) we can find approximate forms of $\phi$, $f$, and $\psi$};
    \node (p8a) [draw, align=center] at (-1, -14.1) {$\phi=\frac{\eta}{\gamma}+\alpha\eta^n$};
    \draw[arrow] (p8) -- (p8a); 
    \node (p8b) [draw, align=center] at (-5, -15) {$\textrm{ln}\left( f\right) =\textrm{ln}\left(\frac{2\gamma}{\gamma+1}\right) - \frac{2\gamma^2 +7\gamma-3}{7-\gamma}\textrm{ln}\left(\frac{\gamma+1-\eta^{n-1}}{\gamma}\right)$};
    \draw[arrow] (p8) -- (p8b);
    \node (p8c) [draw, align=center] at (2.5, -15) {$\textrm{ln} \left( \psi\right) =\textrm{ln}\left(\frac{\gamma+1}{\gamma-1}\right)+\frac{3}{\gamma-1}\textrm{ln}\left( \eta \right) -2\frac{\gamma+5}{7-\gamma}\left(\frac{\gamma+1-\eta^{n-1}}{\gamma}\right)$};
    \draw[arrow] (p8) -- (p8c);
\end{tikzpicture}
\caption{Introduction of non-dimensional pressure, density and velocity along with approximate ratios for large blast radii.} \label{fig:F1}
\end{figure}

While many believe that Taylor used the above approach \cite{RN2}, he instead used first principles and similarity assumptions to arrive at an equation for energy. He eventually arrives at an expression of the same form as \eqref{eq:E} but is able to define the constant $K$. Taylor's process and assumptions are captured in the flow chart in Figure~\ref{fig:F1}.

With this process Taylor used an equation of motion derived from the Navier-Stokes under the assumption of incompressible flow, and no gravitational or viscous effects, leading to the incompressible Euler equations. This process lead him to three equations for the derivatives of the non-dimensionless ratios $f$, $\psi$, and $\phi$ of pressure, density, and velocity respectively. These are expressed as
\begin{equation} \label{eq:f'}
    f'\left(\left(\eta-\phi\right)^2 -\frac{f}{\gamma}\right)=f\left(-3\eta +\phi\left(3+\frac{1}{2}\gamma\right) -2\frac{\gamma\phi^2}{\eta}\right)
\end{equation}
and
\begin{equation} \label{eq:phi'}
    \phi'\left(\eta-\phi\right)=\frac{1}{\gamma}\frac{f'}{\psi}-\frac{3}{2}\phi.
\end{equation}

\footnotetext{Note: There is a typographical error in \citep{RN5}, where Taylor flips this ratio; the end result is the same and this does not lead to any errors.}

\begin{equation} \label{eq:psi'}
    \frac{\psi'}{\psi}=\frac{\phi'+\frac{2\phi}{\eta}}{\eta-\phi}
\end{equation}

\begin{figure}[h]
    \centering
    \includegraphics[width=\textwidth]{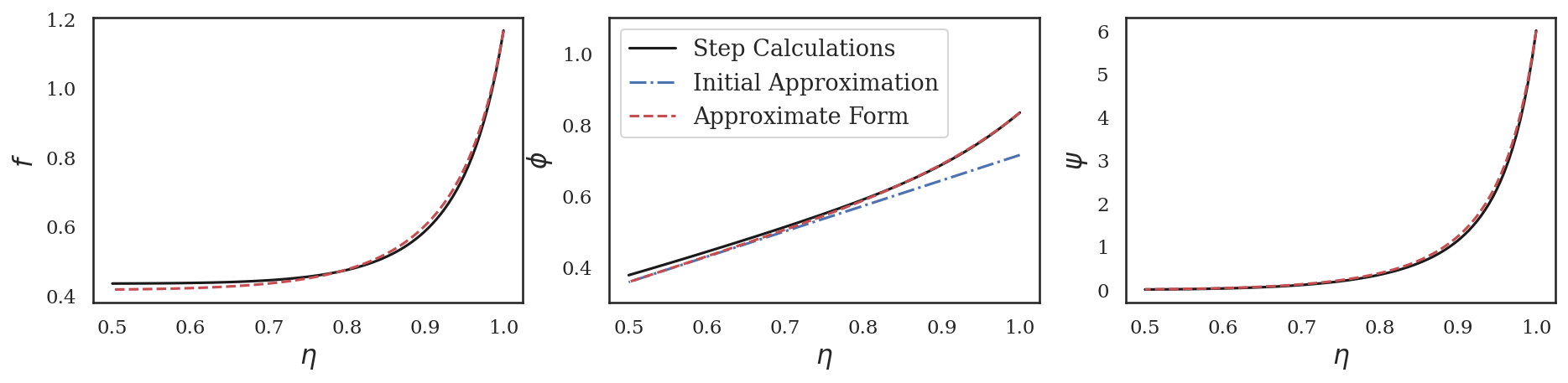}
    \caption{$f$, $\phi$, and $\psi$ vs $\eta$}
    \label{fig:fphipsi}
\end{figure}

From \eqref{eq:f'}, \eqref{eq:phi'}, and \eqref{eq:psi'} Taylor performs a step-by-step calculation for the values of $f$, $\phi$, and $\psi$ for a non-dimensional radius $\eta \in [0,1]$, setting the initial conditions to be at $\eta=1$ and found by the Rankine-Hugonoit solutions, which are calculated in box (7) in Figure \ref{fig:F1}. However, these equations cannot be directly solved unless $f'$ is assumed small. Using that assumption he substitutes equation \ref{eq:psi'} into \ref{eq:f'} and solves for an approximate form of $\phi$. He notes this form, shown in blue in Figure \ref{fig:fphipsi}, does not agree well with the step calculations above $\eta \approx 0.8$, so he suggests the approximate form would not be linear. Instead, the equation would contain a non-linear term of order $n$ which he calculates using the initial conditions set forth by the Rankine-Huginoit solutions at $\eta=1$. From the corrected approximate form of $\phi$ he used equations \eqref{eq:f'}, \eqref{eq:phi'}, and \eqref{eq:psi'} to arrive at approximate forms for $\psi$ and $f$ as well \citep{RN5}, shown in step (8) of Figure \ref{fig:F1}. These step-by-step values as well as the approximated values are shown in Figure \ref{fig:fphipsi}. One can observe that the step-by-step calculations (black) have relatively good agreement with the approximate forms (red), with almost complete agreement for $\psi$. 

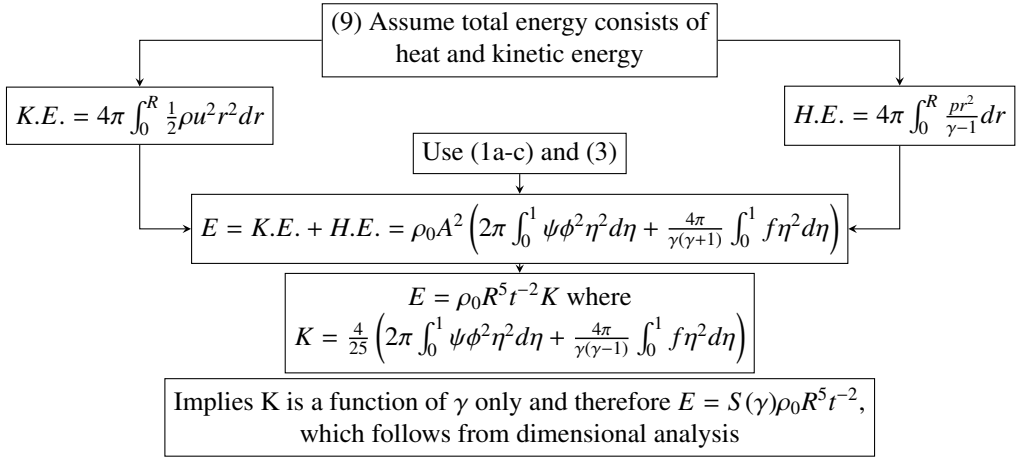
\begin{figure}[h]
    \centering
    \begin{tikzpicture} 
    \node (E) [draw, align=center] at (0, 0) {(9) Assume total energy consists of \\ heat and kinetic energy};
    \node (Ea) [draw, align=center] at (-5, -1) {$K.E. = 4\pi\int_0^R \frac{1}{2}\rho u^2r^2dr$};
    \draw[arrow] (E) -| (Ea);
    \node (Eb) [draw, align=center] at (5, -1) {$H.E. = 4\pi\int_0^R \frac{pr^2}{\gamma-1}dr$};
    \draw[arrow] (E) -| (Eb);
    \node (Ec) [draw, align=center] at (0, -1.5) {Use (1a-c) and (3)};
    \node (Ed) [draw, align=center] at (0, -2.5) {$E=K.E.+H.E. = \rho_0 A^2\left(2\pi\int_0^1 \psi\phi^2\eta^2d\eta  + \frac{4\pi}{\gamma(\gamma+1)}\int_0^1 f\eta^2d\eta\right)$};
    \draw[arrow] (Ea) |- (Ed);
    \draw[arrow] (Eb) |- (Ed);
    \draw[arrow] (Ec) -- (Ed);
    \node (Ee) [draw, align=center] at (0, -3.75) {$E=\rho_0R^5t^{-2}K$ where \\ $K=\frac{4}{25}\left(2\pi\int_0^1 \psi\phi^2\eta^2d\eta  + \frac{4\pi}{\gamma(\gamma-1)}\int_0^1 f\eta^2d\eta\right)$};
    \draw[arrow] (Ed) -- (Ee);
    
    \node (Ef) [draw, align=center] at (0, -5) {Implies K is a function of $\gamma$ only and therefore $E=S(\gamma)\rho_0R^5t^{-2}$, \\ which follows from dimensional analysis};
\end{tikzpicture}
    \caption{Energy calculations}
    \label{fig:energy}
\end{figure}

Once he found these approximate forms he was able to calculate the constant $K$ in his energy equation. In Figure \ref{fig:energy} we show his process for formulating the equation for energy. This marks the end of what he accomplished in his first paper in regards to calculating the energy of the blast \citep{RN5}, as values for radius and time are still needed. Once the Trinity test was declassified he obtained those values and in his second paper \citep{RN6} he linearly interpolates data on the log scale plot of $R^{5/2}$ vs. $t$ and uses the intercept value for $R^5t^{-2}$ in his energy calculations. 

\section{Revisiting Taylor's Analysis}
In this paper we take the same approach as Taylor but make use of modern techniques to improve upon his calculation. We started by reviewing his work as documented in the previous section, to get a full understanding of his approach and assumptions.

\begin{figure}[h]
    \centering
    \includegraphics[width=\textwidth]{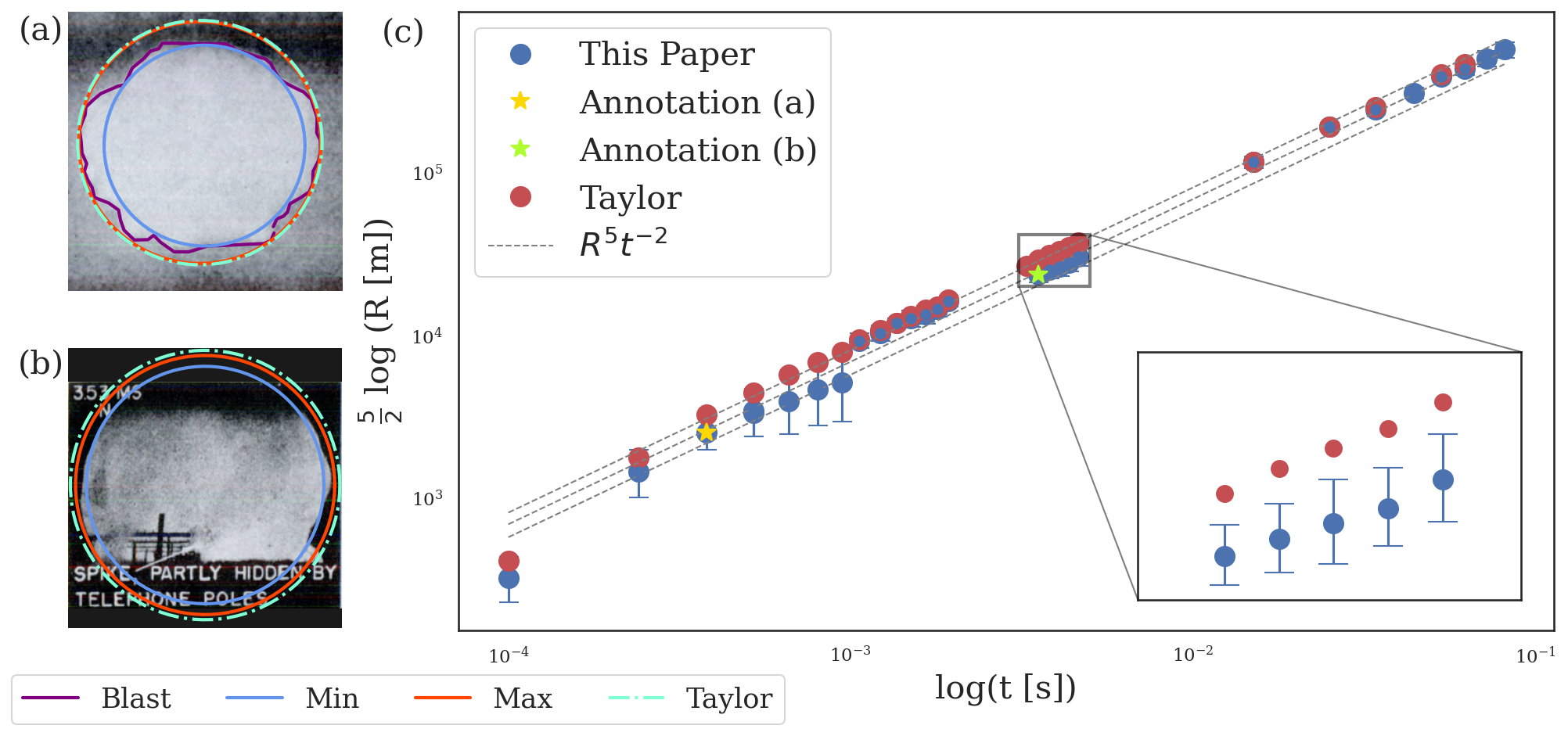}
    \caption{Estimating the blast radius from images taken at different times and plotting the data on a logarithmic scale}
    \label{fig:Rt}
\end{figure}

To emulate his radius measurements we made use of the open source Computer Visualization Annotation Tool (CVAT) \citep{boris_sekachev_2020_4009388} to annotate the sets of photos by Mack \citep{RN4} and the others presented by Taylor in his second paper \citep{RN6}, though we could not find an image for $t=3.26$ ms. With this tool we made an outline of the blast for earlier times as well as annotating two circles that represent the smallest enclosing circle and the largest enclosed circle for all times. For later time stamps, an arrow is provided by Mack \citep{RN4} that designates the shock front, and this is used as the maximum circle for these times. The minimum enclosing circle was checked using the Welzl method \citep{RN10} and the maximum enclosed circle by testing circles centered at Voronoi points \citep{Voronoi1908} of the polygon created by the blast outline. An example of this method along with the radii that Taylor provided for $t=0.36$ and $t=3.53$ ms is shown in figure \ref{fig:Rt} (a) and (b). The larger of the two circles is closer to the approximation that Taylor made, and is the more likely radius as the shock wave is the radius we are interested in, which occurs in front of the fire blast. 

The assumption that Taylor makes based on his first principles analysis, shown in figure \ref{fig:F1} step (3a), is that the radius and time are related by $At=\frac{2}{5}R^{-5/2}$ where A is a constant. Therefore, the radius and time of the blast should correspond to: 

\begin{equation} \label{eq:logRt}
    \frac{5}{2}\textrm{log}(R)= \textrm{log}(t)+\textrm{log}(n)
\end{equation}

Where $n$ is constant and $n^2$ corresponds to the value of $R^5t^{-2}$. Using the radii found in the previous paragraph, we plot the logarithmic-scale graph of $R^{5/2}$ vs $t$ as Taylor did in his second paper \citep{RN6}; this is shown in Figure \ref{fig:Rt}. This figure contains three lines with a slope of one whose y-intercepts correspond to values of $R^5t^{-2}$ for the minimum, maximum, and average radii based on equation \ref{eq:logRt}. These values are calculated using using standard linear least-squares interpolation to fit a linear approximation. Figure \ref{fig:Rt} also displays a zoomed-in view of the calculations from $t \in [3.53, 4.61]$ ms as the difference between our values and Taylor's lead to large discrepancy in the comparison of energy values shown in the next section. 

To calculate $K$ in the energy equation shown in Figure \ref{fig:energy} we used the equations for $\phi$, $\psi$, and $f$ put forth by Taylor in his first paper \citep{RN5} and shown in Figure \ref{fig:F1} step (8). These integrals do not have an analytical solution, so the open-source code  \texttt{equadratures} \citep{RN11} was used to computationally estimate these integrals, i.e.,
\begin{equation}
    K \approx \frac{4\omega_j}{25}\left(2\pi \sum_{j=1}^{J=20} \psi\left( \eta_j \right) \phi^2 \left( \eta_j \right) \eta_j^2   + \frac{4\pi}{\gamma(\gamma-1)} \sum_{j=1}^{J=20} f\left( \eta_j \right) \eta_j^2   \right)
    \label{equ:quadrature}
\end{equation}
where $\left\{\eta_j, \omega_j \right\}_{j=1}^{J=21}$ are Gauss-Legendre quadrature points and weights with order 20 across the support $\eta \in \left[0, 1 \right]$. This order was found to be sufficiently high for numerical precision. We calculated $K$ using \eqref{equ:quadrature} for 10 values of $\gamma$ in the range $\left[ 1.2, 1.667 \right]$. The values calculated with this method show minor differences with those Taylor calculated; see Table~\ref{tab:K-Values}. It is difficult to ascertain what approach Taylor took in calculating the integrals as he simply states in \citep{RN5} that he evaluates the integrals and uses the step-by-step calculations.

\begin{table}[h]
    \centering
    \begin{tabular}{|c|c|c|}
    \hline
        $\gamma$ & This Paper & Taylor \\ \hline 
        \hline
        1.2 & 1.720 & 1.727 \\
        1.3 & 1.145 & 1.167\\
        1.4 & 0.852 & 0.856\\
        1.667 & 0.495 & 0.487\\
        \hline
    \end{tabular}
    \caption{$K$ Values computed using \eqref{equ:quadrature} contrasted with those reported by Taylor}
    \label{tab:K-Values}
\end{table}
Using the values for $R^5t^{-2}$ and K calculated in \eqref{eq:logRt} and \eqref{equ:quadrature} respectively, the total energy was estimated using the same equation as Taylor, shown in figure \ref{fig:energy} step (9). This was done for both the minimum and maximum radii. These values are contrasted to those that Taylor presents in his paper \citep{RN6}. 

\section{Discussion}

\begin{figure}[h]
    \centering
    \includegraphics[width=\textwidth]{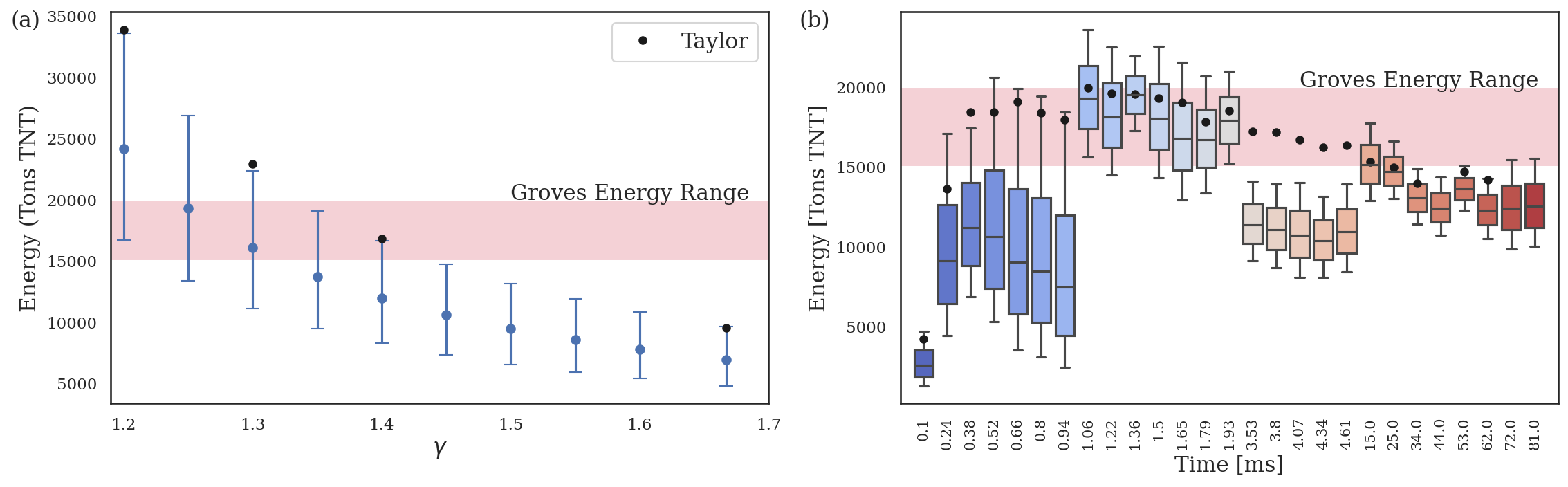}
    \caption{Energy graphs: (left) energy estimates from this paper (blue error bars) across different values of $\gamma$; (right) energy estimates from this paper (box plot) for $\gamma=1.4$ at different times}
    \label{fig:E-graph}
\end{figure}

\begin{figure}[h]
    \centering
    \includegraphics[width=\textwidth]{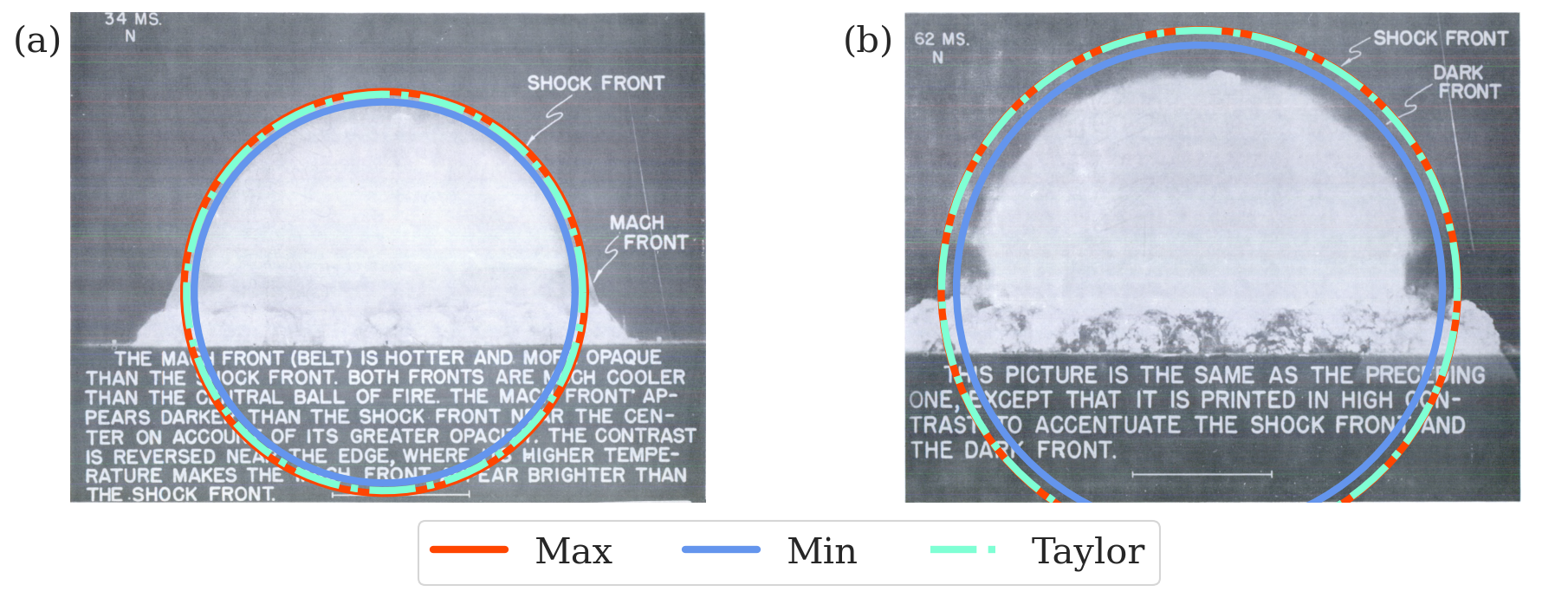}
    \caption{CVAT Annotations Compared With Taylor's Values at Later Times}
    \label{fig:annotations}
\end{figure}

Figure \ref{fig:E-graph} shows the results laid out in the previous section. Figure \ref{fig:E-graph}(a) shows ten energy values at $\gamma \in [1.2, 1.7]$ where the $R^5t^{-2}$ value is taken from the average of the two circles representing the blast, with the blue error bars encompassing the standard deviation. The red shaded rectangle in both (a) and (b) shows the energy range that was given by General Groves in Table~\ref{tab:energy estimates}; his values were used as the range due to the fact that he was directly involved in the Manhattan Project, and therefore the Trinity test. Figure \ref{fig:E-graph}(b) shows the energy value using the $R$, $t$ pairs and the box plot of the energy for those pairs. The scale of the x-axis is not linear and is set to have equal spacing between time measurements to have a better perspective of the box plots. The agreement with the actual energy estimate is best between 1 to 3 ms. This is as expected as the point source solution is invalid at earlier times such as the first time stamp of 0.10 ms. Additionally, the spherical model is no longer valid at later times, approximations at these times can be seen in figure \ref{fig:annotations}. 

From figure \ref{fig:E-graph} we can see that our calculations, specifically for the larger of the two radii, are in very good agreement with those of Taylor. As well as being within the acceptable range of Energy values for $\gamma \approx [1.2, 1.4]$, which Taylor stated was a likely range for the value of $\gamma$. In (b) however between 3.53-4.61 ms the energy values we calculated are different from those calculated using Taylor's radius and time stamps. This is due to differences in the radii measured as these images are unclear and the window is not wide enough to view the whole blast so a certain radius has been assumed, which must have been different for us and Taylor.

Additionally, small differences in radii can lead to large differences in energy as the radii is raised to the 5$^{th}$ power.  Thus, where the radii difference between Taylor and this paper for $t=3.53$ ms differ by only $~2.3 \ m$, the energy difference is around $3000$ tons.

\section{Conclusions}
Taylor's approximation was, for the time and even today, a very good approximation of the energy. He did all this using basic principles and a series of assumptions in order to make all the calculations simpler to compute. The github with all the code/annotation files is published here: \url{https://github.com/03emone/Trinity-Taylor-Files}.

More broadly, the fusion of governing equations with data---inferred from a series of images---for estimating a physical quantity of interest is a task that likely underscores many DCE efforts. Our hope in writing this relatively short paper is that the methods detailed serve as a useful example of how one can: (i) leverage governing equation without necessarily resorting to costly numerical simulations, and (ii) utilize dimensional analysis to gain valueable quantitative insight into a problem.
\bibliographystyle{plainnat}

\end{document}